\documentclass[a4paper]{amsart}
\usepackage{epsfig}
\usepackage{oldlfont}

\newtheorem*{ack}{Acknowledgment}

\newcommand{\bra}[1]{\langle #1 \mid}
\newcommand{\ket}[1]{\mid #1 \rangle}
\newcommand{\braket}[2]{\langle #1 \mid #2 \rangle}

\newcommand{\pic}[5]{\raisebox{#3pt}
{\hspace{#4pt} \epsfig{file=#1.ps,height=#2pt,silent=} 
\hspace{#5pt}}}
\newcommand{\kd}[1]{\mathchoice{
\pic{#1}{24}{-8}{-1}{2}}{
\pic{#1}{11}{-3}{1}{1}}{
\pic{#1}{9}{-2}{-3}{1}}{
\pic{#1}{7}{-1}{-1}{0}}}

\newcommand{\lkd}[1]{\mathchoice{
\pic{#1}{40}{-18}{-1}{2}}{
\pic{#1}{25}{-11}{1}{1}}{
\pic{#1}{20}{-10}{-3}{1}}{
\pic{#1}{18}{-8}{-1}{0}}}

\begin{document}


\title{A Spin Network Primer}

\author{Seth A. Major}

\date{20 August 1999}
\address{ Institut f\"ur Theoretische Physik 
\\ Der Universit\"at Wien \\ Boltzmanngasse 5\\
A-1090 Wien AUSTRIA}
\email{smajor@galileo.thp.univie.ac.at}

\begin{abstract}
Spin networks, essentially labeled graphs, are ``good quantum 
numbers'' for the quantum theory of geometry.  These structures 
encompass a diverse range of techniques which may be used in the 
quantum mechanics of finite dimensional systems, gauge theory, and 
knot theory.  Though accessible to undergraduates, spin network 
techniques are buried in more complicated formulations.  In this paper 
a diagrammatic method, simple but rich, is introduced through an 
association of $2 \times 2$ matrices to diagrams.  This spin network 
diagrammatic method offers new perspectives on the quantum mechanics 
of angular momentum, group theory, knot theory, and even quantum 
geometry.  Examples in each of these areas are discussed.
\end{abstract}

\maketitle

\begin{flushright}
		UWThPh - 1999 - 27
\end{flushright}

\section{Introduction}

Originally introduced as a quantum model of spatial geometry 
\cite{penrose}, spin networks have recently been shown to provide both 
a natural home for geometric operators \cite{RSSpin} and a basis for 
the states of quantum gravity kinematics \cite{BaezSpin}.  At their 
roots, spin networks provide a description of the quantum mechanics of 
two-state systems.  Even with this humble foundation, spin networks 
form a remarkably diverse structure which is useful in knot theory, 
the quantum mechanics of angular momentum, quantum geometry, and other 
areas.

Spin networks are intrinsically accessible to undergraduates, but much 
of the the material is buried in more complex formulations or lies in 
hard-to-find manuscripts.  This article is intended to fill this gap.  
It presents an introduction to the diagrammatic methods of spin 
networks, with an emphasis on applications in quantum mechanics.  In 
so doing, it offers undergraduates not only a fresh perspective on 
angular momentum in quantum mechanics but also a link to leading edge 
research in the study of the Hamiltonian formulation of quantum 
gravity.  One quantum operator of geometry is presented in detail; 
this is the operator which measures the area of a surface.

The history of spin networks goes back to the early seventies when 
Penrose first constructed networks as a fundamentally discrete model 
for three-dimensional space \cite{penrose}.  Difficulties inherent in 
the continuum formulation of physics led Penrose to explore this 
possibility.\footnote{There are more philosophic motivations for this 
model as well.  Mach advocated an interdependence of phenomena: ``The 
physical space I have in mind (which already includes time) is 
therefore nothing but the dependence of the phenomena on one another.  
A completed physics that knew of this dependence would have no need of 
separate concepts of space and time because these would already have 
be encompassed'' \cite{mach} echoing Leibniz's much earlier critique 
of Newton's concept of absolute space and time.  Penrose invokes such 
a Machian principle: A background space on which physical events 
unfold should not play a role; only the relationships of objects to 
each other can have significance \cite{penrose}.} These difficulties 
come from both quantum and gravitational theory as seen from three 
examples: First, while quantum physics is based on noncommuting 
quantities, coordinates of space are commuting numbers, so it appears 
that our usual notion of space conflicts with quantum mechanics.  
Second, on a more pragmatic level, quantum calculations often yield 
divergent answers which grow arbitrarily large as one calculates 
physical quantities on finer and smaller scales.  A good bit of 
machinery in quantum field theory is devoted to regulating and 
renormalizing these divergent quantities.  However, many of these 
difficulties vanish if a smallest size or ``cut-off'' is introduced.  
A discrete structure, such as a lattice, provides such a cut-off.  
Thus, were spacetime built from a lattice or network, then quantum 
field theory would be spared much of the problems of divergences.  
Third, there is a hint coming from general relativity itself.  Since 
regular initial data, say a collapsing shell of matter, can evolve 
into a singularity, relativity demonstrates that the spacetime metric 
is not always well-defined.  This suggests that it is profitable to 
study other methods to model spacetime.  As the absolute space and 
time of Newton is a useful construct to apply in many everyday 
calculations, perhaps continuous spacetime is simply useful as a 
calculational setting for a certain regime of physics.

Motivated by these difficulties, Penrose constructed a discrete model 
of space.  The goal was to build a consistent model from which 
classical, continuum geometry emerged only in a limit.  Together with 
John Moussouris he was able to show that spin networks could reproduce 
the familiar 3 dimensional angles of space -- a ``theory of quantized 
directions'' \cite{JM}.  In this setting, spin networks were trivalent 
graphs labeled by spins.

Later, spin networks were re-discovered by Rovelli and Smolin when 
they were searching for the eigenspace of operators measuring 
geometric quantities such as area and volume \cite{RSSpin}, 
\cite{rsarea}.  In this setting spin networks had to be generalized to 
include graphs with higher valence vertices.  This early work launched 
many studies which resulted in a powerful suite of spin network 
techniques for background-independent quantization.

Spin networks are fantastically useful both as a basis for the states 
of quantum geometry and as a computational tool.  Spin network 
techniques were used to compute the spectrum of area \cite{rsarea} and 
volume \cite{RCgeometry} operators.  Spin networks, first used as a 
combinatorial basis for spacetime, find application in quantum gravity, 
knot theory, and group theory.

This spin network primer begins by associating $2 \times 2$ matrices 
with diagrams.  The first goal is to make the diagrammatics ``planar 
isotopic,'' meaning the diagrams are invariant under smooth 
deformations of lines in the plane.  It is analogous to the 
manipulations which one would expect for ordinary strings on a table.  
Once this is completed, the structure is enriched in Section 
\ref{weavejoin} to allow combinations and intersections between lines.  
This yields a structure which includes the rules of addition of 
angular momentum.  It is further explored in Section 
\ref{angularmomentum} with the diagrammatics of the usual angular 
momentum relations of quantum mechanics.  (A reader more familiar with 
the angular momentum states of quantum mechanics may wish to go 
directly to this section to see how spin networks are employed in this 
setting.)  In Section \ref{grouptheory} this connection to angular 
momentum is used to give a diagrammatic version of the Wigner-Eckart 
theorem.  The article finishes with a discussion on the area operator 
of quantum gravity.

\section{A play on line}

This section begins by building an association between the Kronecker 
delta functions the $2 \times 2$ identity matrix (or $\delta_{A}^{B}$) 
and a line.  It is not hard to ensure that the lines behave like 
elastic strings on a table.  The association and this requirement 
leads to a little bit of knot theory, to the full structure of spin 
networks, and to a diagrammatic method for the quantum mechanics of 
angular momentum.

\subsection{Line, bend and loop}
\label{lineandloop}

The Kronecker $\delta_A^B$ is 
the $2\times 2$ identity matrix in component notation.  Thus,
\[
\left(\delta_{A}^{B} \right) =
\begin{pmatrix}
	1 & 0 \\
	0 & 1
\end{pmatrix}
\]
and $\delta_{0}^{0}=\delta_{1}^{1} =1$ while 
$\delta_{0}^{1}=\delta_{1}^{0} =0$.  
The Latin capital indices, $A$ and $B$ in this expression, 
may take one of two values $0$ or $1$.
The diagrammatics begins by associating the Kronecker $\delta$ to a line
\[
\delta_{A}^{B} \sim \kd{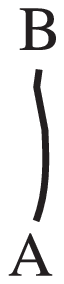}.
\]
The position of the indices on $\delta$ determines the 
location of the labels on the ends of the line.  Applying the 
definitions one has
\[
\kd{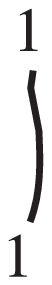} = 1 \text{ and } \kd{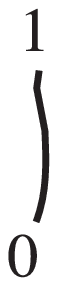} =0.
\]

If a line is the identity then it is reasonable to associate a curve 
to a matrix with two upper (or lower) indices.  There is some freedom 
in the choice of this object.  As a promising possibility, one can 
choose the antisymmetric matrix $\epsilon_{AB}$
\[
\left(\epsilon_{AB} \right) = \left(\epsilon^{AB} \right) =
\begin{pmatrix}
	0 & 1 \\
	-1 & 0
\end{pmatrix}
\]
so that
\[
\epsilon_{AB} \sim \kd{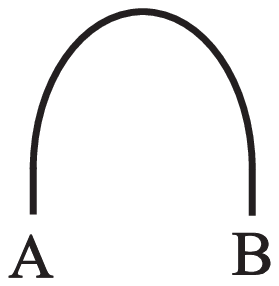}.
\]
Similarly,
\[
\epsilon^{AB} \sim \kd{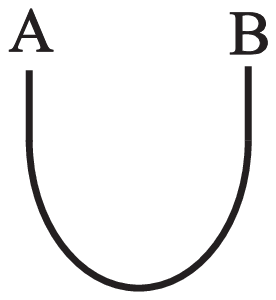}.
\]
As a bent line is a straight line ``with one index lowered'' this 
choice fits well with the diagrammatics: $\delta_{A}^{C} 
\epsilon_{CB} = \epsilon_{AB}$.

After a bit of experimentation with these identifications, 
one discovers two awkward features.  
The diagrams do not match the expected moves of elastic strings in a plane.
First, since $\delta_{A}^{C} 
\epsilon_{CD} \epsilon^{DE} \delta_{E}^{B} = 
\epsilon_{AD}\epsilon^{DB} = - \delta_{A}^{B}$, straightening a line 
yields a negative sign:
\begin{equation}
	\label{wiggleminus}
\kd{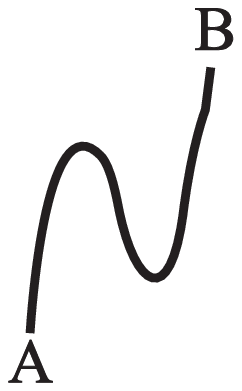} = - \kd{lineAB}.
\end{equation}
Second, as a consequence of $\epsilon_{AD} \epsilon_{BC} \epsilon^{CD}
= - \epsilon_{AB}$,
\begin{equation}
	\label{bentminus}
\kd{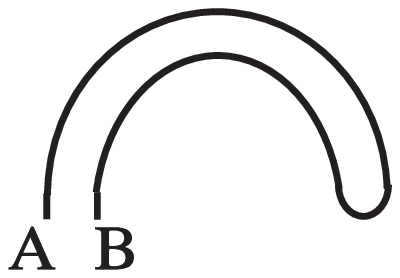} = - \kd{capAB}.
\end{equation}
However, these ``topological''  
difficulties are fixed by modifying the definition of a bent 
line. One can add an $i$ to the antisymmetric tensors
\[
\epsilon_{AB} \rightarrow \tilde{\epsilon}_{AB} = i \epsilon_{AB}
\text{  so that  } \tilde{\epsilon}_{AB} = \kd{capAB}.
\]
Since each of the two awkward features contains a pair of 
$\epsilon$'s the $i$ fixes these sign problems.  
However, there is one more property to investigate.

On account of the relation $\delta_{A}^{D} \delta_{B}^{C} 
\tilde{\epsilon}_{CD}
= - \tilde{\epsilon}_{AB}$ one has (The indices $C$ and $D$ are added 
to the diagram for clarity.)
\[
\kd{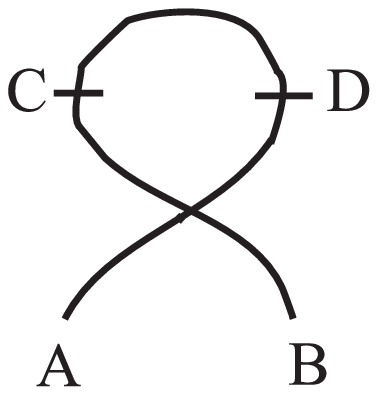} = - \kd{capAB}
\]
--  not what one would expect for strings. 
This final problem can be cured by associating a minus sign to each
crossing.

Thus, by associating an $i$ to every $\epsilon$ and a sign to 
every crossing, the diagrams behave as continuously deformed 
lines in a plane \cite{penrose}.  
The more precise name of this concept is known as 
planar isotopy.  Structures which can be moved about in this way are 
called topological.  What this association of curves to $\delta$'s
and $\tilde{\epsilon}$'s accomplishes is that it allows one to perform algebraic 
calculations by moving lines in a plane.
 
A number of properties follow from the above definitions.  The value of a 
simple closed loop takes a negative value\footnote{This led 
Penrose to dub these ``negative dimensional tensors'' \cite{penrose}.  
In general relativity, the dimension of a space is given by the trace of 
the metric, $g_{\mu \nu} g^{\mu \nu}$, hence the name.}
\begin{equation}
	\label{loop}
	\kd{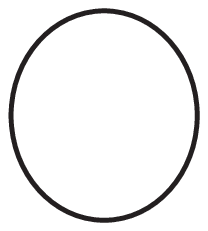} = -2,
\end{equation}
since $\tilde{\epsilon}_{AB}\, \tilde{\epsilon}^{AB} = - \epsilon_{AB}
\, \epsilon^{AB} =-2$;
a closed line is a number.  This turns out to be a generic result in that 
a spin network which has no open lines is equivalent to a 
number.

A surprisingly rich structure emerges when crossings,
are considered.  For instance
the identity, often called the ``spinor identity,'' links a pair of epsilons 
to products of deltas
\[
\epsilon_{AC} \, \epsilon^{BD} = \delta_{A}^{B} \, \delta_{C}^{D} - 
\delta_{A}^{D} \, \delta_{C}^{B}.
\]
Using the definitions of the $\tilde{\epsilon}$ matrices one may show 
that, diagrammatically, this becomes
\begin{equation}
	\label{binorid}
	\pic{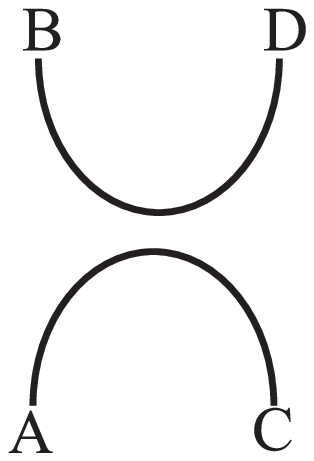}{30}{-12}{-1}{2} + \pic{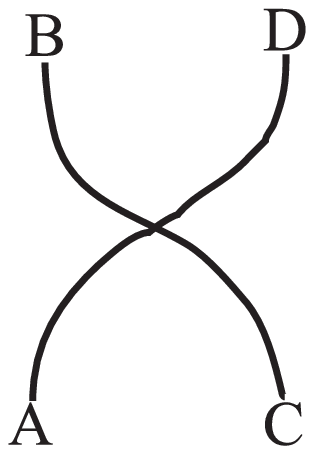}{30}{-12}{-1}{2} +
	\pic{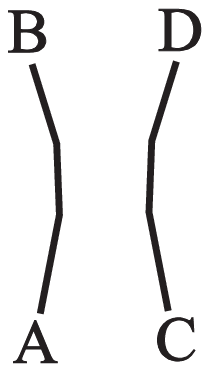}{30}{-12}{-1}{2} = 0.
\end{equation}
Note that the sign changes, e.g. $ - \delta_{A}^{D} \, \delta_{C}^{B}$ becomes
$+ \kd{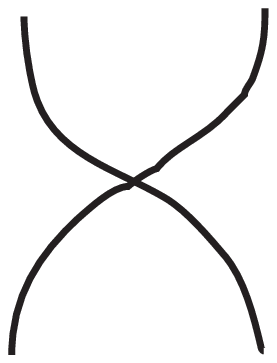}$.  This diagrammatic relation of Eq. (\ref{binorid}) is known as 
``skein relations'' or the ``binor identity.''  The
utility of the relation becomes evident when one realizes that the
equation may be applied anywhere within a larger diagram.

One can also decorate the structure by ``weighting'' or ``tagging''
edges.\footnote{There is some redundancy in notation.  Numerical
(or more general) labels associated to edges are frequently called
weights or labels.  The term ``tag'' encompasses the meaning of 
these labels as well as operators on lines.}
Instead of confining the diagrams to simply be a sum of products of 
$\delta$'s and $\epsilon$'s, one can include other objects with a tag.
For instance,  one can associate a tagged line to any $2 \times 2$ 
matrix such as $\psi_{A}^{B}$
\[
\psi_{A}^{B} \sim \pic{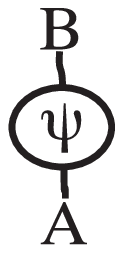}{26}{-10}{-1}{2}.
\]
These tags prove to be useful notation for angular momentum operators 
and for the spin networks of quantum geometry.
Objects with only one index, can frequently be represented as
Kronecker delta functions with only one index.  For example,
\[
u^{A} = \kd{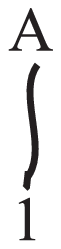}.
\]

The result of these associations is a topological structure in which algebraic 
manipulations of $\delta$'s, $\epsilon$'s, and other $2 \times 2$ 
matrices are
encoded in manipulations  of open or closed lines.  For instance, straightening a 
wiggle is the same as simplifying a product of two 
$\tilde{\epsilon}$'s to a single $\delta$.
It also turns out that the algebra is ``topological:'' 
Any two equivalent algebraic expressions are represented by two 
diagrams which can be continuously transformed into each other. 
Making use of a result of Reidemeister and the identities above it is a
few lines of $\delta  \epsilon$-algebra to show that the spin network
diagrammatics is topologically invariant in a plane.

\subsection{Reidemeister Moves}

Remarkably, a knot\footnote{A mathematical knot is a knotted, closed loop. 
One often encounters collections of, possibly knotted, knots.  These 
are called links.} in three dimensional space can be
continuously deformed into another knot, if and only if, the planar
projection of the knots can be transformed into each other via a sequence of 
four moves called the ``Reidemeister moves''  \cite{R}.  Though the 
topic of this primer is mainly on two dimensional diagrams, the  Reidemeister moves
are given here in their full generality -- as projections of knots in three 
dimensional space.  While in two dimensions one has only an intersection, 
$\kd{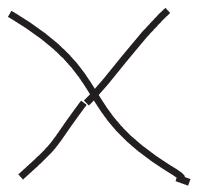}$, 
when two lines cross, in three dimensions one has the ``over 
crossing,'' $\kd{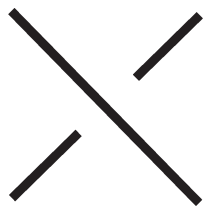}$ and the ``undercross,'' $\kd{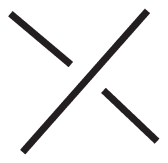}$, as 
well as the intersection $\kd{cross}$.

There are four moves:
\begin{itemize}
\item {\bf Move 0:} In the plane of projection, one can make smooth 
deformations of the curve
\[
\kd{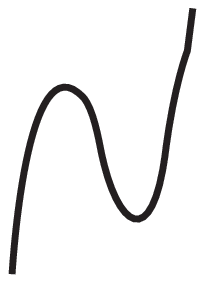} \sim \kd{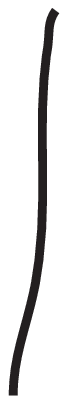}.
\]
\item {\bf Move I:} As these moves are designed for one dimensional
objects, a curl may be undone
\[
\kd{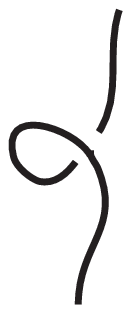} \sim \kd{line}.
\]
This move does not work on garden-variety string.  The 
string becomes twisted (or untwisted).  (In fact, this is the way yarn 
is made.)
\item {\bf Move II:} The overlaps of distinct curves are not knotted
\[
\kd{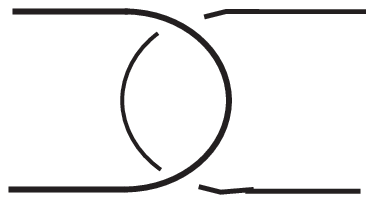} \sim \kd{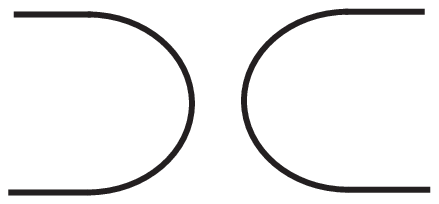}.
\]
\item {\bf Move III:} One can perform planar deformations under (or 
over) a diagram 
\[
\kd{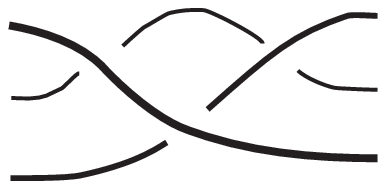} \sim \kd{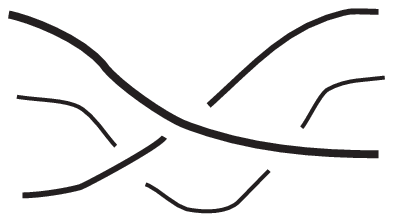}.
\]
\end{itemize}
With a finite sequence of these moves the projection of a knot may be
transformed into the projection of any other knot which is 
topologically equivalent to
the original. If two knots may be expressed as the other with a 
sequence of these moves then the knots are called ``isotopic.'' 
Planar isotopy is generated by all four moves
with the significant caveat that there are no crossings
$\kd{crossovr}$, only intersections $\kd{intersect}$.  Planar
isotopy may be summarized as the manipulations one would expect for 
elastic, non-sticky strings on a table top -- if they are infinitely thin.

Move I on real strings introduces a twist in the string.  This move is violated by 
any line which has some spatial extent in the transverse direction 
such as ribbons.  Happily, there are diagrammatic spin networks for these 
``ribbons'' as well \cite{knotsandphysics}, \cite{KL}.

\subsection{Weaving and joining}
\label{weavejoin}

The skein relations of Eq. (\ref{binorid}) show 
that, given a pair of lines there is one linear relation among the three quantities:  
$\kd{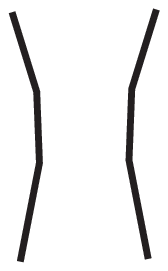}$, $\kd{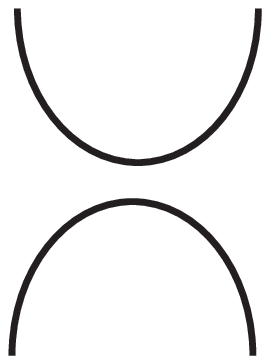}$, and $\kd{cross}$.  So a set of graphs 
may satisfy many linear relations. It would be nice to select a basis   
which is independent of this identity.  After some work, this may be 
accomplished by choosing the antisymmetric combinations of
the lines -- ``weaving with a 
sign.''\footnote{Note that, because of the additional sign associated 
to crossings, the ``antisymmetrizer'' symmetrizes the indices in the 
$\delta \epsilon$ world.} 
The simplest example is for two lines
\begin{equation}
\label{2line}
\kd{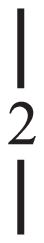} = \frac{1}{2} \left( \kd{lines} - \kd{cross} \right).
\end{equation}
For more than two lines the idea is the same. One sums over permutations of the
lines, adding  a sign for each crossing.  The general definition is
\begin{equation}
	\label{asym}
\kd{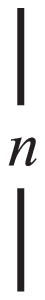} := \frac{1}{n!} \sum_{\sigma \in S_{n}} (-1)^{|\sigma|} \, \kd{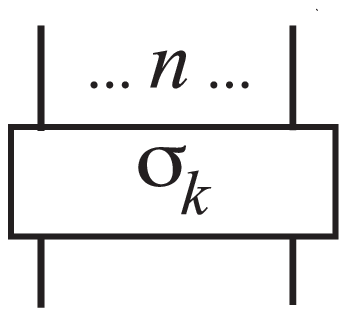}
\end{equation}
in which a $\sigma$ represents one permutation of the $n$ 
lines and $|\sigma|$ is the minimum number of crossings for this permutation.
The boxed $\sigma$ in the diagram represents the action of the 
permutation on the lines.  It can be drawn by writing $1 \, 2\,  \dots 
\, n$,  then permutation just above it, and connecting the same elements 
by lines.

In this definition, the label $n$ superimposed on the edge record
the number of ``strands'' in the edge. Edge are usually labeled this 
way, though I will leave simple 1-lines
unlabeled.  Two other notations are used for this weaving with a sign
\[
\kd{linen} = \kd{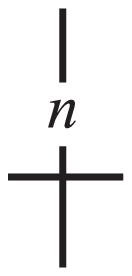} = \kd{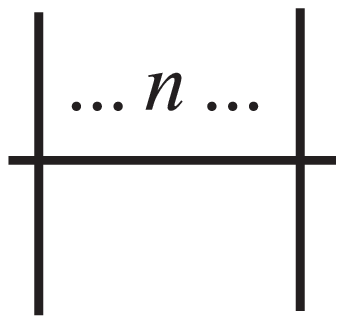}.
\]

These antisymmetrizers have a couple of lovely properties,
retacing and projection:
The antisymmetrizers are ``irreducible,'' or vanish when a pair of lines is 
retraced
\begin{equation}
	\label{irreduce}
\kd{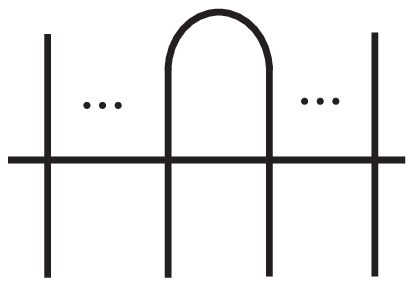} = 0.
\end{equation}
which follows from the antisymmetry.  Using this and the binor 
identity of Eq. (\ref{binorid}) one may show that the antisymmetrizers are 
``projectors'' (the combination of two is equal to one)
\[
\kd{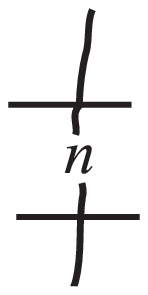} = \kd{linen}.
\]

Making the simplest closed diagram out of these lines gives
the loop value often denoted as $\Delta_{n}$
\[
\kd{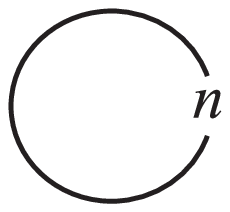} = \Delta_{n} = (-1)^{n}(n+1).
\]
The factor $n+1$ expresses the ``multiplicity'' of the 
number of possible ``$A$-values'' on an edge with $n$ strands.
Each line in the edge carries an
index, which takes two possible values.  To see this note that for an edge with 
$a$ strands the sum of the indices $A, B, C, ...$ is $0,1,2, ... ,a$.
So that the sum takes $a+1$ possible values. One may show using the 
recursion relations for $\Delta_{n}$\footnote{The loop value 
satisfies $\Delta_{0} =1, \Delta_{1} = -2$, and $\Delta_{n+2} = (-2) 
\Delta_{n+1} - \Delta_{n}$.} that the loop value is equal to 
this multiplicity.
As we will see in Section \ref{angularmomentum} the number of
possible combinations is the dimension of the 
representation.

As an example of the loop value, the 2-loop, 
has value 3.  This is easily checked using the 
relations for the basic loop value (Eq. (\ref{loop})) and the
expansion of the 2-line using the skein relation
\begin{equation}
	\label{2temper}
	\kd{line2} = \kd{lines} + \tfrac{1}{2} \kd{cupcap}.
\end{equation}

Edges may be further joined into networks by making use of 
internal trivalent vertices
\[
\lkd{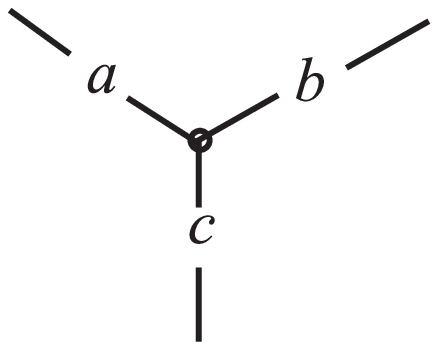} =  \lkd{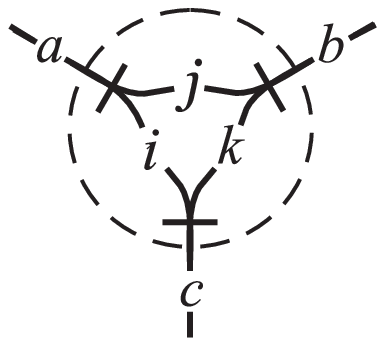}.
\]
The dashed circle is a magnification of the dot in the diagram on the 
left.  Such dashed curves indicate spin network structure at a point.  
The ``internal'' labels $i,j,k$ are positive integers determined
by the external labels $a,b,c$ via
\[
i = (a + c - b)/2, \ \ j = ( b+c- a)/2, \ \ { \rm and} \ \ k = (a+b - 
c)/2.
\]
As in quantum mechanics the external labels must satisfy the triangle inequalities
\[
a + b \geq c, \ \ b + c \geq a,  \ \ a + c \geq b
\]
and the sum $a+b+c$ is an even integer.  The necessity of these relations can be seen 
by drawing the strands through the vertex.

With this vertex one can construct many more 
complex networks.  After the loop, the next simplest closed graph has 
two vertices,
\[
\theta(a,b,c) = \kd{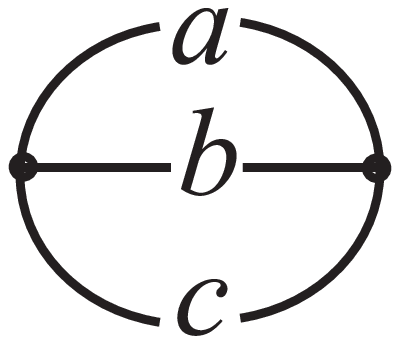}.
\]
The general evaluation, given in the appendix, of this diagram 
is significantly more complicated.
As an example I give the evaluation of $\theta(1,2,1)$
using Eq. (\ref{2temper}),
\begin{equation*}
	\begin{split}
		\kd{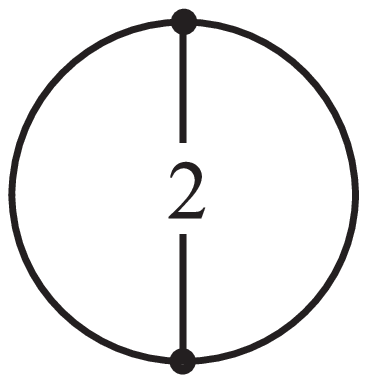} &= \kd{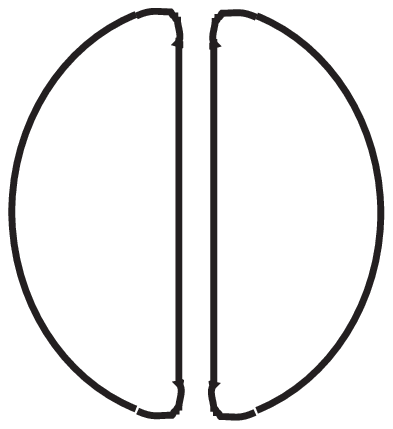} + \frac{1}{2} \kd{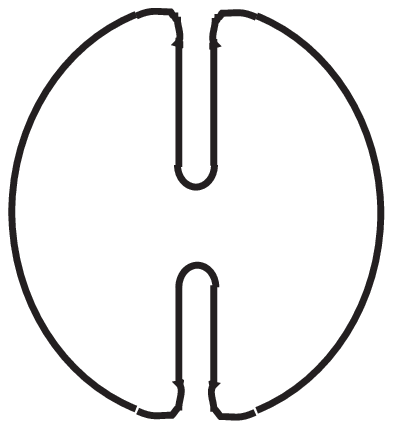}\\
		&= (-2)^{2} + \frac{1}{2}(-2) = 3.
	\end{split}
\end{equation*}
One can build ever more complicated networks. In fact, one can soon 
land a dizzying array of networks.  I have collected a small zoo in the 
appendix with full definitions.

Now all the elements are in place for the definition of spin 
networks.  A spin network consist of a graph, with edges and vertices,
and labels. The labels, associated edges, represent 
the number of strands woven into edges.  Any vertex with more 
than three incident edges must also be labeled to specify a 
decomposition into trivalent vertices.  The graphs of spin networks 
need not be confined to a plane.  In a projection of a spin network embedded in 
space, the crossings which appear in the projection may be shown as in 
the Reidemeister moves with over-crossing ``$\kd{crossovr}$''  and 
under-crossing ``$\kd{crossudr}$''.

\section{Angular momentum representation}
\label{angularmomentum}

As spin networks are woven from strands which take 
two values, it is well-suited to represent two-state systems. 
It is perhaps not 
surprising that the diagrammatics of spin networks include the 
familiar $\ket{j m}$ representation of angular momentum.
The notations are related as
\begin{eqnarray*}
\ket{ \tfrac{1}{2} \ \tfrac{1}{2} } = u^A \sim \kd{uA} \text{  and} \\
\ket{ \tfrac{1}{2} \  - \tfrac{1}{2} } = d^A \sim \kd{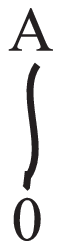}.
\end{eqnarray*}
(Secretly, the ``$u$'' for ``up'' tell us that the index $A$ only takes 
the value 1. Likewise  ``$d$'' tells us the index is 0.)  The inner product is
given by linking upper and lower indices, for instance
\[
\braket{ \tfrac{1}{2} \ \tfrac{1}{2} } {\tfrac{1}{2} \ \tfrac{1}{2}}
\sim \kd{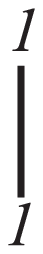} = 1.
\]
For higher representations \cite{JM}
\begin{equation}
\label{jline}
\ket{j \  m} := \ket{r \, s} = N_{rs} \; 
\underbrace{u^{(A} u^B \dots u^C}_{r} \; 
\underbrace{d^D d^E \dots d^{F)}}_{s}
\end{equation}
in which
\begin{equation}
\label{rsdefg}
N_{rs} = \left( \frac{ 1 }{ r! \, s! \, (r+s)! }\right)^{1/2}, \ \ j = 
\tfrac{ r+s}{2}, \ \ { \rm and} \ \  m = \tfrac{ r-s} {2}.
\end{equation}
The parentheses in Eq. (\ref{jline}) around the indices indicate 
symmetrization, e.g. $u^{(A}d^{B)} = u^{A}d^{B}+u^{B}d^{A}$.  
The normalization $N_{rs}$ ensures that the states are orthonormal in
the usual inner product.  A useful representation of this state is in 
terms of the trivalent vertex.  Using the notation ``$\kd{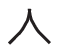}$'' for 
$u$ and similarly for  $d$ I have
\[
\ket{j \, m} \sim \lkd{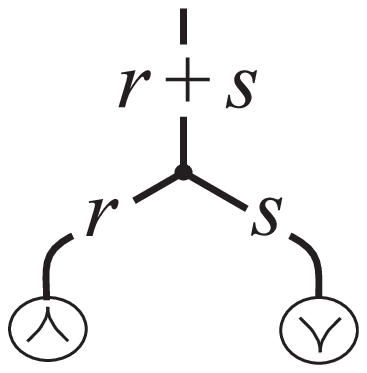}.
\]

Angular momentum operators also take a diagrammatic form.    
As all spin networks are built from spin-$\tfrac{1}{2}$
states, it is worth exploring this territory first.
Spin-$\tfrac{1}{2}$ operators have a representation in terms of 
the Pauli matrices
\[
\sigma_{1} = \begin{pmatrix}
	0 & 1 \\
	1 & 0 
\end{pmatrix}
,
\;
\sigma_{2}=\begin{pmatrix}
	0 & -i \\
	i & 0
\end{pmatrix}
, \;
\sigma_{3} = \begin{pmatrix}
	1 &0 \\
	0 & -1
\end{pmatrix}
\]
with
\[
\hat{S}_{i} = \frac{\hbar}{2} \sigma_{i}
\]
for $i=1,2,3$. One has
\[
\frac{\sigma_{3}}{2} \ket{\tfrac{1}{2} \tfrac{1}{2}} 
= \tfrac{1}{2} \ket{\tfrac{1}{2} \tfrac{1}{2}},
\]
which is expressed diagrammatically as 
\[
\kd{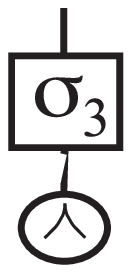} = \tfrac{1}{2} \kd{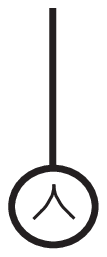}.
\]
Or, since Pauli matrices are traceless,
\[
\kd{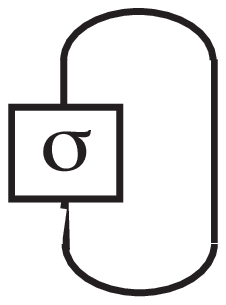}=0,
\]
and using Eq (\ref{2temper}) one has \cite{roberto}
\[
\kd{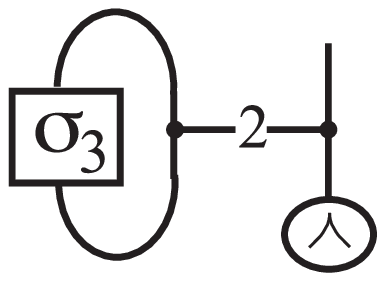}= \tfrac{1}{2} \kd{line+}.
\]
A similar relation holds for the states 
$\ket{\tfrac{1}{2} \, - \tfrac{1}{2}}$.  The basic action 
of the spin operators can be described as a ``hand'' which acts on the
state by ``grasping'' a line \cite{RS}.  The result, after using
the diagrammatic algebra, is either a multiple of the 
same state, as for $\sigma_{3}$, or a 
new state.  If the operator acts on more than one line, a higher dimensional 
representation, then the total action is the sum of the graspings on each
edge.\footnote{This may be shown by noticing that
\[
\kd{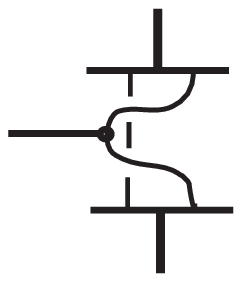} = \kd{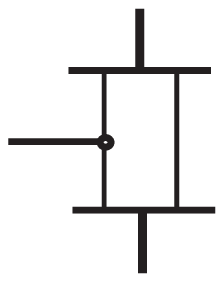}
\text{ ,  so that }
\kd{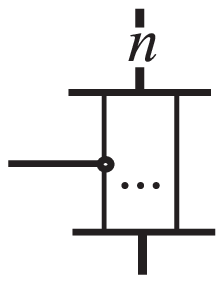} + \kd{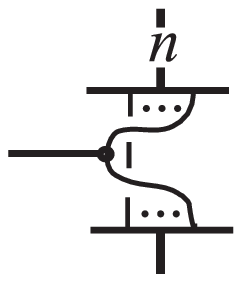} + \dots = n \kd{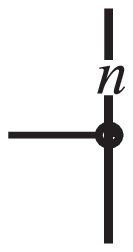},
\]
as may be derived using Eqs. (\ref{binorid}) and (\ref{irreduce}).}

The $\hat{J}_{z}$ operator can be constructed out of the $\sigma_{3}$
matrix.  The total angular momentum 
$z$-component is the sum of individual measurements on each of the 
sub-systems.\footnote{The operator is $\hat{J}_{z}= \hbar 
\sum_{i=1}^{2j+1} 1 \otimes ... \otimes \left( 
\tfrac{\sigma_{3}}{2}\right)_{i}  \otimes
... \otimes 1$ where the sum is over the possible positions of 
the Pauli matrix.}  In diagrams, the action of the $\hat{J}_{z}$ operator 
becomes
\begin{equation*}
	\begin{split}
\hat{J}_{z} \ket{j \, m} &\sim \kd{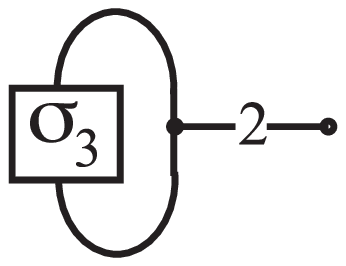} \lkd{trirs} \\
		                 &= r \lkd{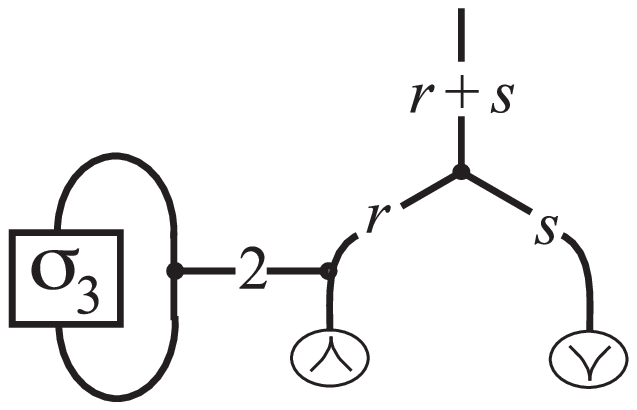} + s \lkd{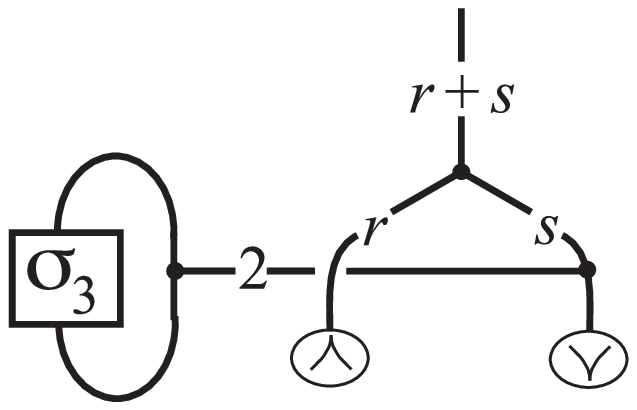} \\
		                 &= \hbar \left( \frac{r}{2} - \frac{s}{2} \right) 
		                 \lkd{trirs} \\
		                 &\equiv \hbar \, m \ket{j \, m}.
	\end{split}
\end{equation*}
The definition of the quantities $r$ and $s$ was used in the last line.

This same procedure works for the other angular momentum operators as 
well.  The $\hat{J}_{x}$ operator is constructed from the Pauli
matrix $\sigma_{1}$.  When acting on one line the operator $\hat{J}_{x}$
matrix ``flips the spin'' and leaves a factor
\[
\kd{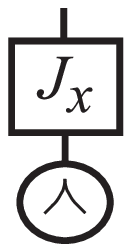} = \hbar \tfrac{1}{2} \kd{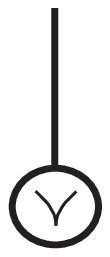}.
\]
The reader is encouraged to try the same procedure for 
$\hat{J}_{y}$.  

The raising and lowering operators are constructed
with these diagrams as in the usual algebra.  For the raising 
operator $\hat{J}_{+} = \hat{J}_{1} +i \hat{J}_{2}$ one has
\[
\hat{J}_{+} \ket{j \, m} \sim \hbar s \lkd{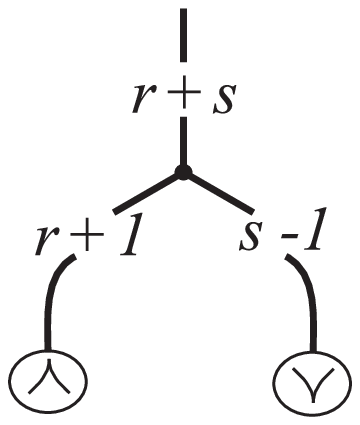}.
\]
In a similar way one can compute
\[
\hat{J}_{-} \hat{J}_{+} \ket{j \, m} = \hbar^{2} (r+1) s \ket{j \, m}
\]
from which one can compute the normalization of these operators:  Taking 
the inner product with $\bra{j \, m}$ gives 
the usual normalization for the raising operator
\[
\hat{J}_{+} \ket{j \, m} = \hbar \sqrt{s(r+1)} \ket{j \, m} 
= \hbar \sqrt{(j-m)(j+m+1)} \ket{j \, m}.
\]
Note that since $r$ and $s$ are non-negative and no larger than $2j$, the 
usual condition on $m$, $-j \leq m \leq j$, is automatically satisfied.

Though a bit more involved, the same procedure goes through for the 
$\hat{J}^{2}$ operator.  It is 
built from the sum of products of operators $\hat{J}^{2} = 
\hat{J_{x}}^{2} +\hat{J_{y}}^{2}+\hat{J_{z}}^{2}$.  Acting once with the 
appropriate Pauli operators, one finds
\begin{equation*}
\begin{split}
\hat{J}^{2} \ket{ j \, m} &\sim \hbar \frac{\hat{\sigma}_{1}}{2} \left[
\frac{r}{2} \lkd{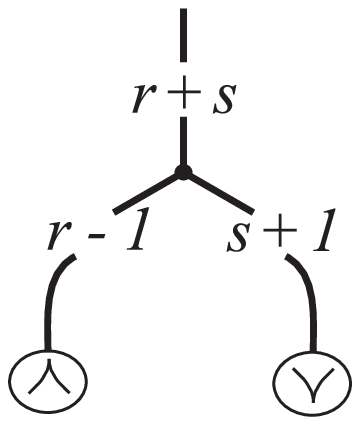} + \frac{s}{2} \lkd{trir+1s-1} \right]
+ \hbar \frac{\hat{\sigma}_{2}}{2} \left[
\frac{i r}{2} \lkd{trir-1s+1} - \frac{i s}{2} \lkd{trir+1s-1} 
\right]  \\
&+ \hbar \frac{\hat{\sigma}_{3}}{2} \frac{(r+s)}{2} \lkd{trirs}.
\end{split}
\end{equation*}
Acting once again, some happy cancellation occurs and the result is
\[
\hat{J}^{2} \ket{ j \, m} = \frac{\hbar^{2}}{2} \left( \frac{r^{2} +s^{2}}{2} 
+rs +r +s \right) \ket{j \, m}
\]
which equals the familiar $j(j+1)$.  Actually, there is a pretty 
identity which gives another route to this result.  The Pauli
matrices satisfy \cite{roberto}
\begin{equation}
\label{pauli}
\frac{1}{2} \sum_{i=1}^{3} \sigma_{i \, A}^{B} \, \sigma_{i \, C}^{D}
= \kd{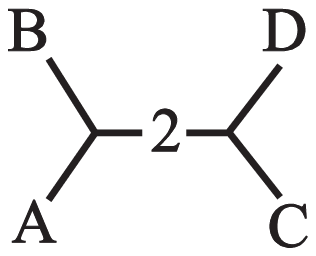}
\end{equation}
so the product is a $2$-line. Similarly, the $\hat{J}^{2}$ operator may be
expressed as a $2$-line. As will be shown in Section \ref{areaop}
this simplifies the above calculation considerably.

\section{A bit of group theory}
\label{grouptheory}

As we have seen, spin networks, inspired by expressing simple
$\delta$ and $\epsilon$ matrices in terms of diagrams, are closely 
related to the familiar angular momentum representation of quantum 
mechanics.   This section makes a brief excursion into group theory to 
exhibit two results which take a clear diagrammatic form, 
Schur's Lemma and the Wigner-Eckart Theorem. 

Readers with experience with some group theory may have noticed 
that spin network edges 
are closely related to the irreducible representations of $SU(2)$.  
The key difference is that, on account of the sign conventions chosen
in Section \ref{lineandloop}, the usual symmetrization of representations 
is 
replaced by the antisymmetrization of Eq. (\ref{asym}).  In fact, 
each edge of the spin network is an irreducible representation.  
The tags on the edges can identify how these are generated --  
through the spatial dependence of a phase, for instance.

Since this diagrammatic algebra is designed to handle the combinations
of irreducible representations, all the familiar results of
representation theory have a diagrammatic form.  For instance, Schur's
Lemma states that any matrix $T$ which commutes with two
(inequivalent) irreducible representations $D_g$ and $D'_g$ of
dimensions $a+1$ and $b+1$ is either zero or a multiple of the identity
matrix
\[
T D_g = D'_g T \ \text{ for all } \ g \in G \implies T = \begin{cases}
0 & {\rm if } \ \ a \neq b \\ \lambda & { \rm if } \ \ a=b \end{cases}.
\]
Diagrammatically, this is represented as
\[
\lkd{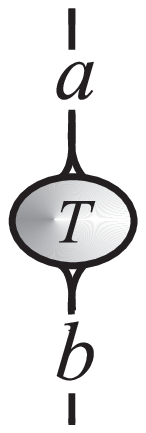}= \lambda \  \delta_a^b \ 
\kd{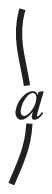},
\text{ where  }
\lambda = \frac{ \lkd{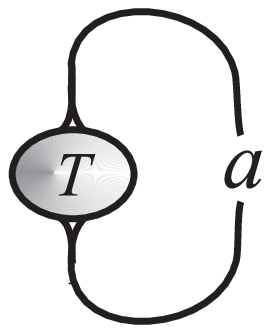}}{\Delta_{a}}.
\]
The constant of proportionality is given by $\lambda$ which, being a 
closed diagram, evaluates to a number.

The Wigner-Eckart theorem also takes a nice form in the diagrammatic 
language, providing an 
intuitive and fresh perspective on the theorem.  It
can help those who feel lost in the mire of 
irreducible tensor operators, reduced matrix elements, and 
Clebsch-Gordon coefficients.
A general operator $T^{j}_{m}$ grasping a line in the $j_{1}$ 
representation ($2j_{1}$ lines) to give a $j_2$ representation is 
expressed as
\[
\pic{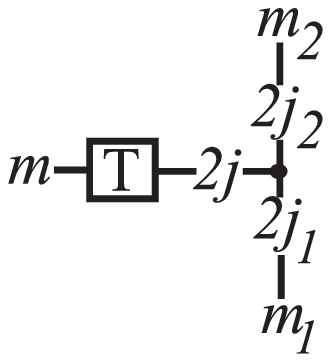}{48}{-20}{-1}{2} \sim \bra{ j_2 m_2} T^{j}_{m} \ket{j_{1} m_{1}}.
\]
Just from this diagram and
the properties of the trivalent vertex, it is already clear that
\[
|j_{1}-j_{2}| \leq j \leq j_{1} +j_{2}.
\]
Likewise it is also the case that 
\[
m_{2} = m_{1} + m.
\]
These results are the useful ``selection rules'' which are often given 
as a corollary to the Wigner-Eckart theorem.
Notice that the operator expression is a 
diagram with the three legs $j$, $j_{1}$, and $j_{2}$.
This suggests that it might be possible to express the operator as a 
multiple of the basic trivalent vertex.\footnote{Since the 
Clebsch-Gordan symbols are complete, any map from $a \otimes b$
to $c$ must be a multiple of the Clebsch-Gordan or $3j$-symbol.}  
Defining
\[
\pic{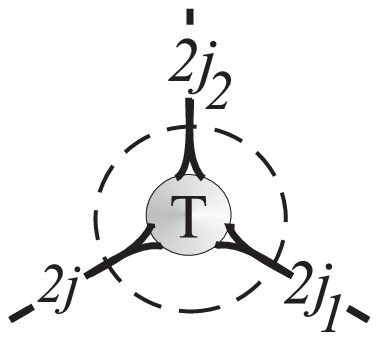}{48}{-20}{-1}{2} := \pic{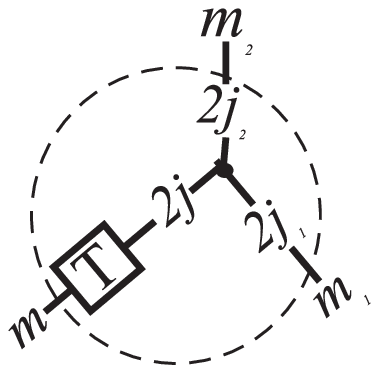}{48}{-20}{-1}{2},
\]
one can combine the two lower legs together with Eq. (\ref{join}).  
Applying Schur's Lemma, one finds
\begin{equation}
\label{WED}
\lkd{opT2} = \sum_{c=|j-j_{1}|}^{j+j_{1}} \frac{\Delta_{c}}{\theta(2j, 
2j_{1}, c)} \lkd{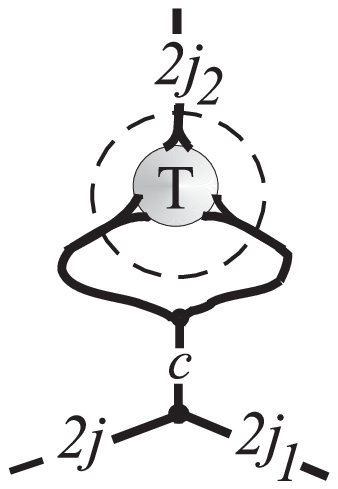} = \omega
\kd{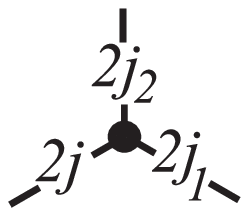},
\end{equation}
where
\[
\omega = \frac{1}{\theta(2j,2j_{1}, 2j_{2})} \lkd{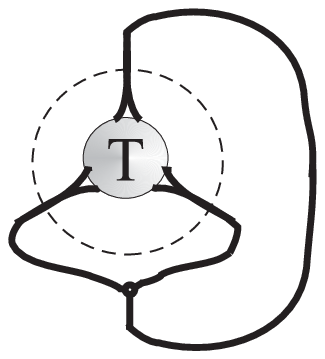}.
\]
This relation expresses the 
operator in terms of a multiple of the trivalent 
vertex.  It also gives a computable expression of the multiplicative 
factor.
Comparing the first and last terms with the usual form of the 
theorem --
\footnote{See, for instance, Reference \cite{mess} pg. 573 or 
Reference \cite{jones} pg. 116.}
\[
\bra{j_{2} m_{2}} T^{j}_{m} \ket{j_{1} m_{1}} = \bra{j_{2}} 
|T^{j}_{m}| \ket{j_{1}} \braket{j m j_{1} m_{1}}{j_{2} m_{2}}  
\]
-- one can immediately see that the reduced matrix element $\bra{j_{2}} 
|T^{j}_{m}| \ket{j_{1}}$ is the $\omega$ of Eq. (\ref{WED}).
In this manner, any invariant tensor may be represented as a
labeled, trivalent graph.

\section{Quantum Geometry: Area operator}
\label{areaop}

In this final example of the spin network diagrammatic algebra, the 
spectrum of the area operator of quantum gravity is derived.  Before 
beginning, I ought to remark that the hard work of defining what is 
meant by the quantum area operator is not done here.  The presentation 
instead concentrates on the calculation of the spectrum.

There are many approaches to constructing a quantum theory of gravity.  
The plethora of ideas arises in part from the lack of experimental 
guidance and in part from the completely new setting of general 
relativity for the techniques of quantization.  One promising 
direction arises out of an effort to construct a 
background-independent theory which meets the requirements of quantum 
mechanics.  This field may be called ``loop quantum gravity'' or 
``spinet gravity.''  (See Ref.  \cite{CRrev} for a recent review.)  
The key idea in this approach is to lay aside the perturbative methods 
usually employed and, instead, directly quantize the Hamiltonian 
theory.  Recently this field has bloomed.  There is now a 
mathematically rigorous understanding of the kinematics of the theory 
and a number of (in principle testable) predictions of quantum 
geometry.  One of the intriguing results of this study of quantum 
geometry is the discrete nature of space.

In general relativity the degrees of freedom are encoded in the metric 
on spacetime.  However, it is quite useful to use new variables to 
quantize the theory \cite{AA}.  Instead of a metric, in the canonical 
approach the variables are an ``electric field,'' which is the 
``square root'' of the spatial metric, and a vector potential.  The 
electric field ${\bf E}$ is not only vector but also takes $2\times2$ 
matrix values in an ``internal'' space.  This electric field is 
closely related to the coordinate transformation from curved to flat 
coordinates (a triad).  The canonically conjugate ${\bf A}$, usually 
taken to be the configuration variable, is similar to the electric 
vector potential but is more appropriately called a ``matrix 
potential'' for ${\bf A}$ also is matrix valued.  It determines the 
effects of geometry on spin-$\tfrac{1}{2}$ particles as they are moved 
through space.\footnote{For those readers familiar with general 
relativity the potential determines the parallel transport of 
spin-$\frac{1}{2}$ particles.} (See Refs.  \cite{AArev} and 
\cite{CRrev} for more on the new variables.)  States of loop quantum 
gravity are functions of the potential ${\bf A}$.  A convenient basis 
is built from kets $\ket{s}$ labeled by spin networks $s$.  In this 
application of spin networks, they have special tags or weights on the 
edges of the graph.  Every strand $e$ of the gravitational spin 
network has the ``phase'' associated to it.\footnote{In more detail, 
every edge has a ``holonomy.''  or path ordered exponential (i.e. 
${\cal P} \exp \int_{e} dt \, {\bf \dot{e}}(t) \cdot {\bf A} (e(t)) $) 
associated to it.  See, for example, Ref.  \cite{AArev}.} An 
orientation along every edge helps to determine these phases or 
weights.  The states of quantum geometry are encoded in the 
knottedness and connectivity of the spin networks.

In classically gravity the area of a surface $S$ is the integral
\[
A_{S} = \int_{S} d^{2}x \sqrt{g},
\]
in which $g$ is the determinant of the metric on the
surface.\footnote{The flavor of such an
additional dependence is already familiar in
flat space intergrals in spherical coordinates:
\[
A = \int r^{2} \sin(\theta) d\phi d\theta.
\]}
The calculation simplifies if the surface is specified by $z=0$ 
in an adapted coordinate system.
Expressed in terms of ${\bf E}$, the area of a surface $S$ 
only depends on the $z$-vector component \cite{rsarea} - \cite{flrarea}
\begin{equation}
	\label{areae}
	A_{S} = \int_{S} d^{2}x \sqrt{E_{z} \cdot E_{z}}.
\end{equation}
The dot product is in the ``internal'' space.  It is the same product 
between Pauli matrices as appears in Eq.  (\ref{pauli}).  In the spin 
network basis ${\bf E}$ is the momentum operator.  As $p \rightarrow 
-i \hbar \tfrac{d}{dx}$ in quantum mechanics, the electric field 
analogously becomes a ``hand,'' ${\bf E} \rightarrow -i \hbar \kappa 
\kd{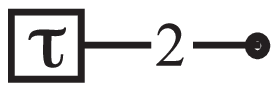}$.  The $\tau$ is proportional to a Pauli matrix, $\tau = 
\tfrac{i}{2} \sigma$.  The $\kappa$ factor is a sign: It is positive 
when the orientations on the edge and surface are the same, negative 
when the edge is oriented oppositely from the surface, and vanishes 
when the edges is tangent to the surface.  The ${\bf E}$ operator acts 
like the angular momentum operator $\hat{J}$.  Since the ${\bf E}$ 
operator vanishes unless it grasps an edge, the operator only acts 
where the spin network intersects the surface.

The square of the area operator is calculated first.  Calling the 
square of the integrand of Eq.  (\ref{areae}) $\hat{O}$, the 
two-handed operator at one intersection is
\begin{equation}
	\hat{O} \ket{s} =  - \sum_{e_{I}, e_{J}} \kappa_{I} \, \kappa_J \,  
	\hat{J}_{I} \cdot
	\hat{J}_{J} \ket{s}
\end{equation}
where the sum is over edges $e_{I}$ at the intersection.  Here, 
$\hat{J}_I$ denotes the vector operator $\hat{J} = \hat{J}_x + 
\hat{J}_y + \hat{J}_z $ acting on the edge $e_{I}$.  This $\hat{O}$ is 
almost $\hat{J}^{2}$ but for the sign factors $\kappa_{I}$.  The area 
operator is the sum over contributions from all parts of the spin 
network which thread through the surface.  In terms of $\hat{O}$ over 
all intersections $i$
\begin{equation*}
	\hat{A}_{S} \ket{s} = \frac{G}{4 c^{3}} \sum_{i} 
	\hat{O_{i}}^{1/2} \ket{s},
\end{equation*}
including the dimensional constants.

\begin{figure} 
\begin{center}
\begin{tabular*}{\textwidth}{c@{\extracolsep{\fill}}
c@{\extracolsep{\fill}}}
\pic{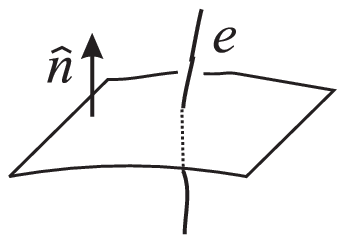}{50}{-15}{-1}{2}
 & 
\pic{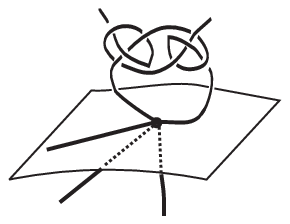}{50}{-15}{-1}{2}\\
(a.) &  (b.) 
\end{tabular*}
\end{center}
\label{figure} \caption{ Two types of intersections of a spin network 
with a surface (a.)  One isolated edge $e$ intersects the surface 
transversely.  The normal $\hat{n}$ is also shown.  (b.)  One vertex 
of a spin network lies in the surface.  All the non-tangent edges 
contribute to the area.  Note that the network can be knotted.}
\end{figure}

As a first step, one can calculate the action of the operator 
$\hat{O}$ on an edge $e$ labeled by $n$ as depicted in Figure 
\ref{figure}(a.).  In this case, the hands act on the same edge so the 
sign is $1$, $\kappa_I^2 =1$, and the angle operator squared becomes 
proportional to $\hat{J}^2$!  In the calculation one may make use of 
the Pauli matrix identity of Eq.  (\ref{pauli})
\begin{equation*}
	\begin{split}
		\hat{O}_{e} \ket{s}
		&=  - \hat{J}^2 \ket{s} \\
		&= - \hbar^2 \, \frac{n^{2}}{2} \, \kd{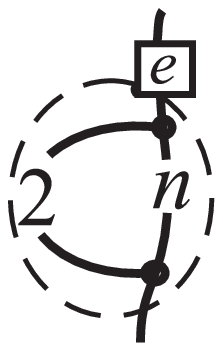} \ket{(s-e)}.
	\end{split}
\end{equation*}
The edge is shown in the the diagram so it is removed  
spin network $s$ giving the state $\ket{(s-e)}$.  
Now the diagram may be reduced using the recoupling 
identities. The bubble may be extracted with Eq. (\ref{bubbleident})
\begin{equation*}
	\begin{split}
		\hat{O}_{e} \ket{s}
		&=- \hbar^2 \, \frac{n^{2}}{2} \, \kd{opJJ} \ket{(s-e)} \\
		&= - \hbar^2 \, \frac{n^{2}}{2} \, 
		\frac{\theta(n,n,2)}{\Delta_{n}} \, \kd{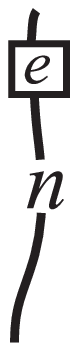} 
		\ket{(s-e)} \\
		& = - \hbar^2 \, \frac{n^{2}}{2} \, \left( - 
		\frac{n+2}{2n} \right) \ket{s} \\
		& = \hbar^2 \, \frac{n(n+2)}{4} \ket{s},
	\end{split}
\end{equation*}
in which Eq. (\ref{tnn2}) was also used in the second line.  Putting 
this result into the area 
operator, one learns that the area coming from all the transverse 
edges is \cite{rsarea}
\begin{equation}
	\begin{split}	
	\hat{A}_{S} \ket{s} &= \frac{G\hbar}{c^{3}} \sum_{i}
	\sqrt{ \frac{n_{i}(n_{i}+2)}{4} } \ket{s} \\
	&= l_{P}^{2} \sum_{i} \sqrt{j_{i}(j_{i}+1)} \ket{s}.
	\end{split}
\end{equation}
The units $\hbar, c,$ and $G$ are collected into the Planck length $l_{P} 
= 
\sqrt{\tfrac{G \hbar} {c^{3}}} \sim 10^{-35} \, m$.   The result is 
also re-expressed in terms of the 
more familiar angular momentum variables $j=\tfrac{n}{2}$.

The full spectrum of the area operator is found by considering all the 
intersections of the spin network with the surface $S$ including 
vertices which lie on the surface as in Figure \ref{figure}(b.).  
Summing over all contributions \cite{alarea}
\[
\hat{A}_{S} \ket{s} = \frac{l_{P}^{2}}{2} \sum_{v} \left[
2 j_{v}^{u}(j_{v}^{u} +1) + 2 j_{v}^{d}(j_{v}^{d} +1) -
j_{v}^{t}(j_{v}^{t} +1) \right]^{1/2} \ket{s}
\]
in which $j_{v}^{u}$ ($j_{v}^{d}$) is the total spin with a positive 
(negative) sign $\kappa$ and $j_{v}^{t}$ is the total spin of edges 
tangent to the surface at the vertex $v$.

This result is utterly remarkable in that the calculation predicts 
that space is discrete.  Measurements of area can only take these 
quantized values.  As is the case in many quantum systems there is a 
``jump'' from the lowest possible non-zero value.  This ``area 
quanta'' is $\tfrac{\sqrt{3}}{4} l_{p}^{2}$.  In an analogous fashion, 
as for an electron in a hydrogen atom, surfaces make a quantum jump 
between states in the spectrum of the area operator.

\section{Summary}

This introduction to spin networks diagrammatics offers a view of the 
diversity of this structure.  Touching on knot theory, group theory, 
and quantum gravity this review gives a glimpse of the applications.  
These techniques also offer a new perspective on familiar angular 
momentum representations of undergraduate quantum mechanics.  As shown 
with the area operator in the last section, it is these same 
techniques which are a focus of frontier research in the Hamiltonian 
quantization of the gravitational field.

\begin{ack} 
It is a pleasure to thank Franz Hinterleitner and Johnathan Thornburg 
for comments on a draft of the primer.  I gratefully acknowledge 
support of the FWF through a Lise Meitner Fellowship.
\end{ack}

\appendix
\section{Loops, Thetas, Tets and all that}

This appendix contains the basic definitions and formulae of 
diagrammatic recoupling theory using the conventions of Kauffman and 
Lins \cite{KL} -- a book written in the context of the more general 
Temperley-Lieb algebra .

The function $\theta(m, n, l)$ is given by
\begin{equation} \label{theta}
\theta(m,n,l)= \kd{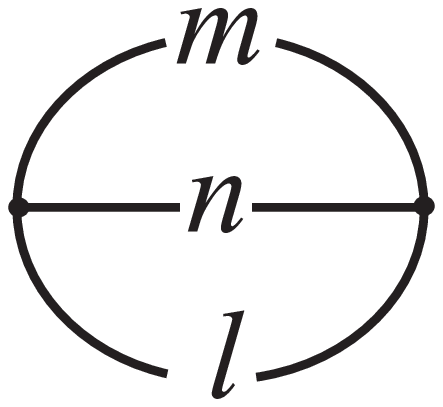} =
(-1)^{(a+b+c)}{(a+b+c+1)!a!b!c! \over (a+b)!(b+c)!
(a+c)!}
\end{equation}
where $a=(l+m-n)/2$, $b=(m+n-l)/2$, and $c=(n+l-m)/2$.
An evaluation which is useful in calculating the spectrum of the area
operator is 
$\theta(n,n,2)$, for which $a=1$, $b=n-1$, and $c=1$.
\begin{equation}
	\label{tnn2}
	\theta(n,n,2) = (-1)^{(n+1)} \frac{(n+2)! \, (n-1)!}{(2 n!)^{2}}
	= (-1)^{(n+1)} \frac{(n+2) (n+1)} {2n}.
\end{equation}

A ``bubble'' diagram is proportional to a single edge.
\begin{equation}
	\label{bubbleident}
\pic{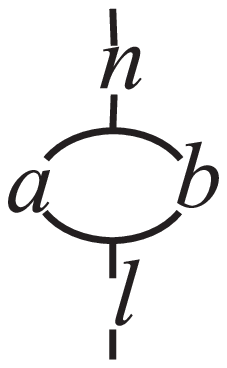}{30}{-12}{-1}{2} 
= \delta_{nl}\frac{ (-1)^{n} \theta(a, b, n)}{(n+1) }
\pic{linen}{30}{-12}{-1}{2}.
\end{equation}

The basic recoupling identity relates the
different ways in which three angular momenta, say $a$, $b$,
and $c$, can couple to form a fourth one, $d$. The two possible
recouplings are related by
\begin{equation}
\label{recoupling}
\kd{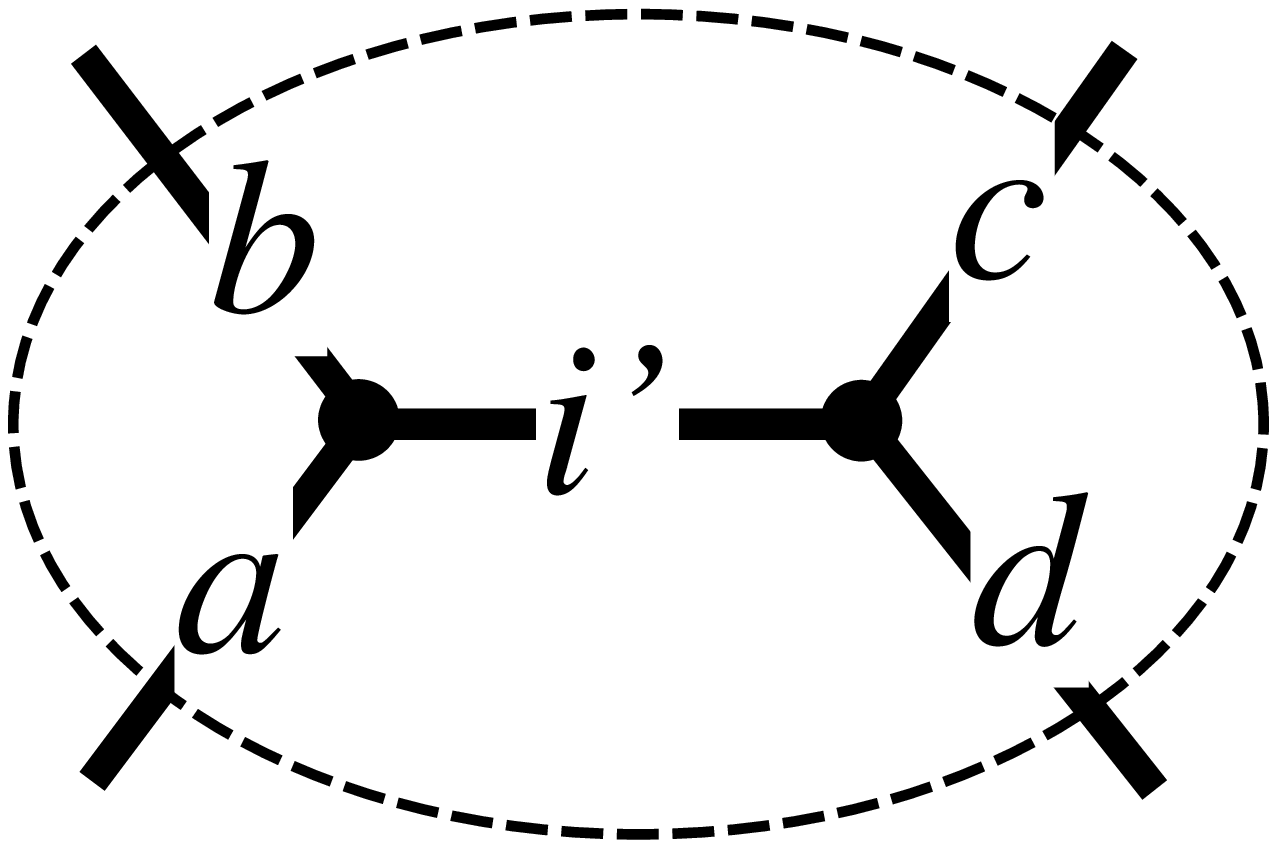} = \sum_{|a-b| \leq i \leq(a+b)}
\left\{ \begin{array}{ccc} a & b & i \\  c & d & i' 
\end{array} \right\} \kd{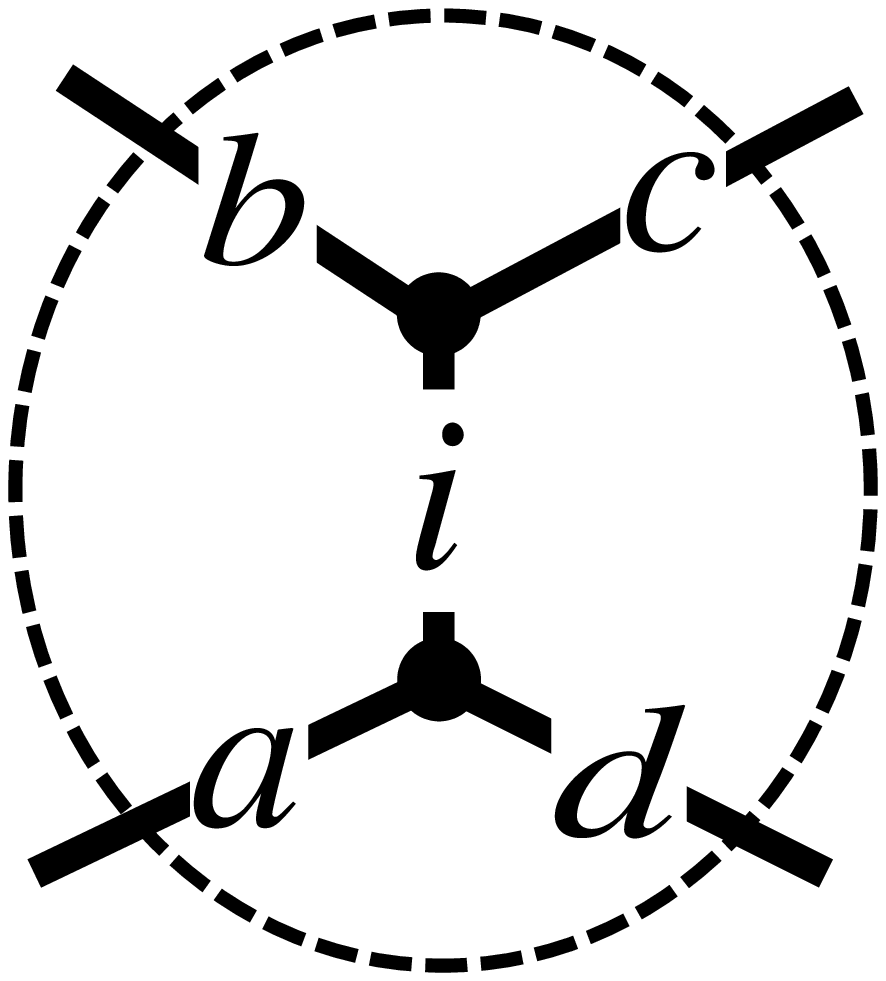}
\end{equation}
where on the right hand side is the $6j$-symbol defined below.
It is closely related to the $Tet$ symbol. This is defined by \cite{KL}
\begin{equation}
\begin{split}
\label{TetDef}
\lkd{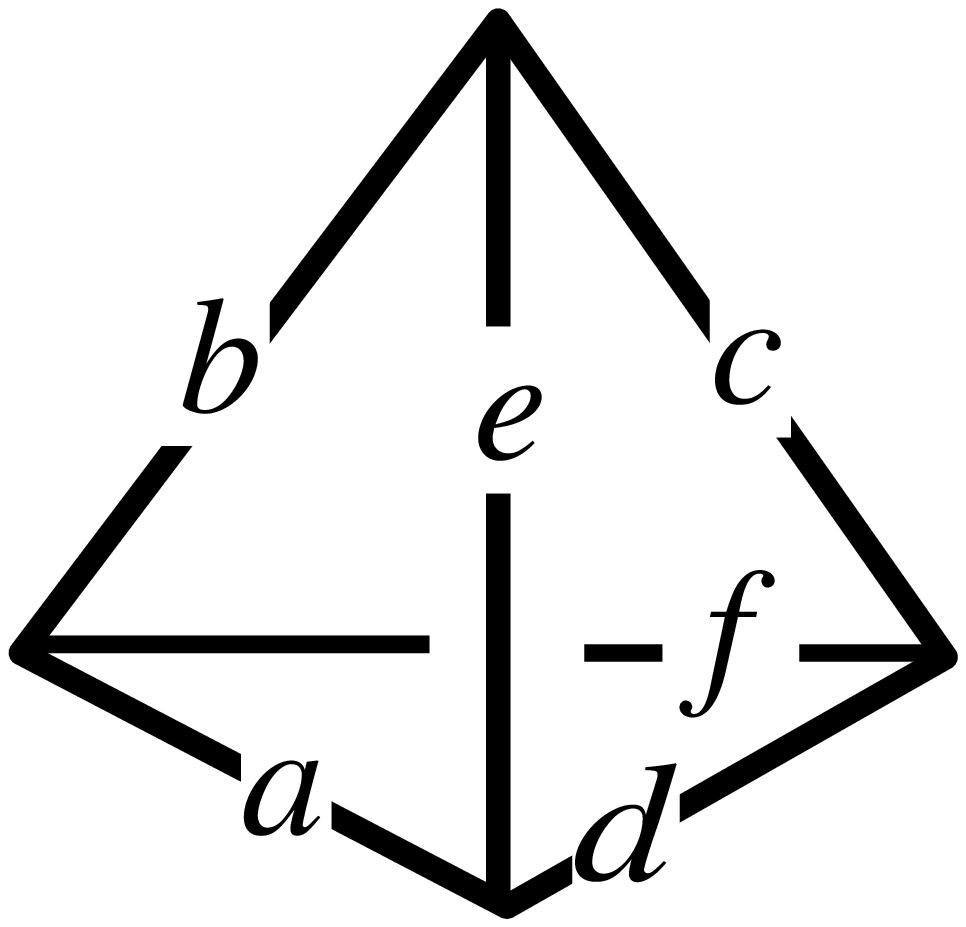} &= \lkd{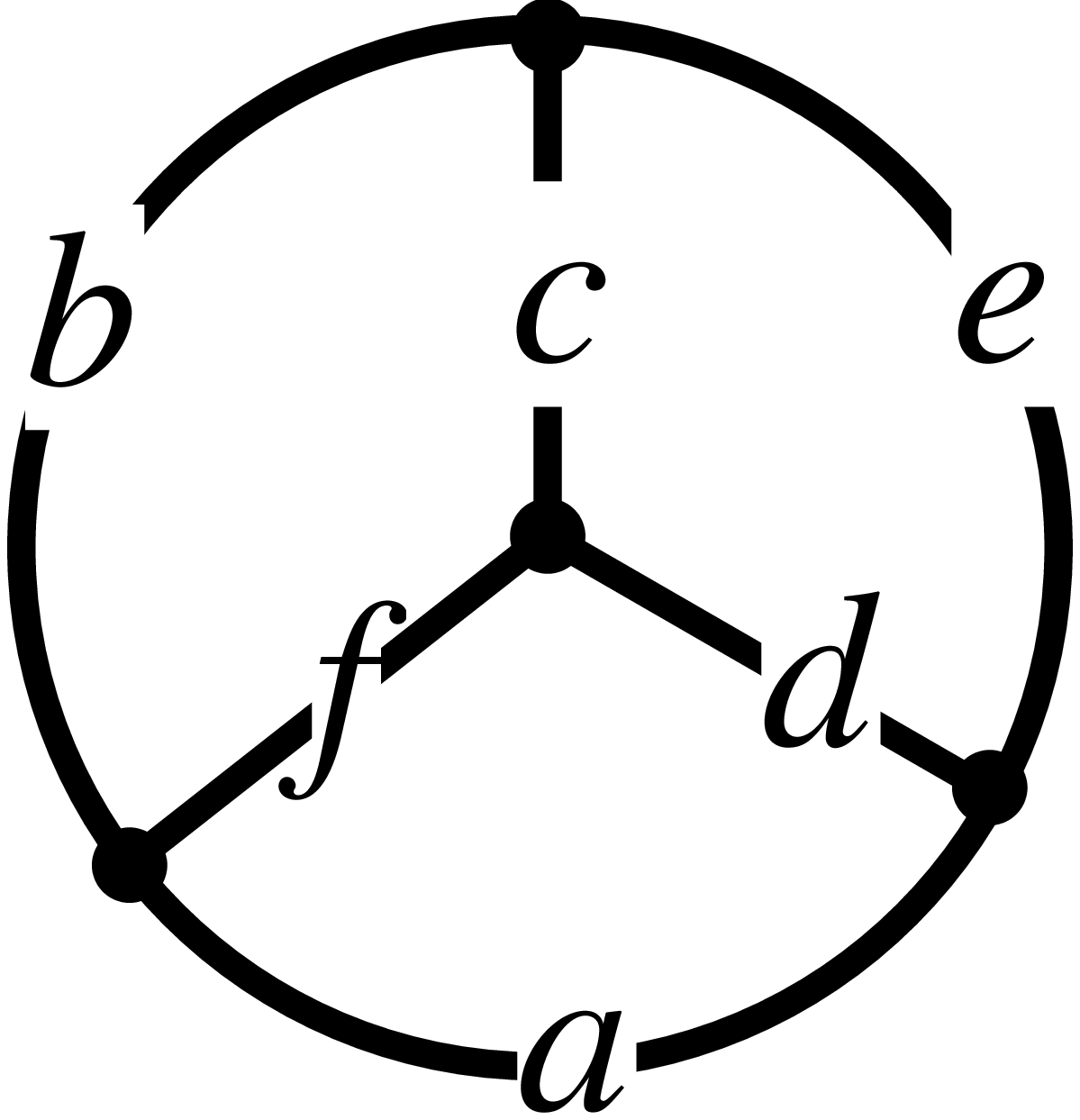} = Tet 
\begin{bmatrix} a & b & e \\ c & d & f \end{bmatrix} \\
Tet \begin{bmatrix} a & b & e \\ c & d & f \end{bmatrix} &= N 
\sum_{m \leq s \leq S} (-1)^s  {  (s+1)! \over
\prod_i\, (s-a_i)! \; \prod_j \, (b_j -s)! } \\
N &= { \prod_{i,j}\, [b_j - a_i]! \over a!b!c!d!e!f!}
\end{split} \end{equation}
in which
\begin{equation} \begin{align}
a_1 &= \tfrac{1}{2} ( a +d + e) & b_1 &= \tfrac{1}{2} ( b +d + e+ f) 
\nonumber \\
a_2 &= \tfrac{1}{2} ( b +c + e) & b_2 &= \tfrac{1}{2} ( a +c + e +f) 
\nonumber \\
a_3 &= \tfrac{1}{2} ( a +b + f) & b_3 &= \tfrac{1}{2} ( a +b + c+d) 
\nonumber \\
a_4 &= \tfrac{1}{2} ( c +d + f) & m={\rm max}\, \{a_i\} \ \ 
M={\rm min}\, \{b_j\} \nonumber
\end{align} \end{equation}

The $6j$-symbol is then defined as
\begin{equation*}
\left\{ \begin{array}{ccc} a & b & i \\ c & d & j \end{array} \right\}
:=
{ Tet \begin{bmatrix} a & b & i \\ c & d & j \end{bmatrix} \Delta_i
\over \theta(a,d,i) \; \theta(b,c,i) }.
\end{equation*}
These satisfy a number of properties including the orthogonal identity
\begin{equation*}
	\sum_{l}
	\left\{ \begin{array}{ccc} a & b & l \\ c & d & j \end{array} \right\}
	\,
	\left\{ \begin{array}{ccc} d & a & i \\ b & c & l \end{array} \right\}
	= \delta_{i}^{j}	
\end{equation*}
and the Biedenharn-Elliot or Pentagon identity
\begin{equation*}
	\sum_{l}
	\left\{ \begin{array}{ccc} d & i & l \\ e & m & c \end{array} \right\}
	\left\{ \begin{array}{ccc} a & b & f \\ e & l & i \end{array} \right\}
	\left\{ \begin{array}{ccc} a & f & k \\ d & d & l \end{array} \right\}
	=
	\left\{ \begin{array}{ccc} a & b & k \\ c & d & i \end{array} \right\}
	\left\{ \begin{array}{ccc} k & b & f \\ e & m & c \end{array} \right\}.
\end{equation*}

Two lines may be joined via
\begin{equation}
	\label{join}
	\kd{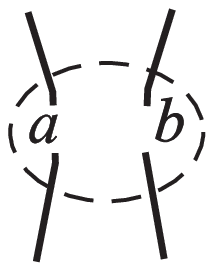} = \sum_{c} \frac{\Delta_{c}}{\theta(a,b,c)} \kd{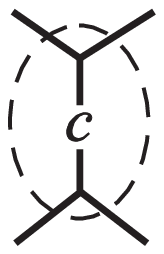}.
\end{equation}

One also has occasion to use the coefficient of the ``$\lambda$-move''
\begin{equation} \begin{split} \label{lmove}
\kd{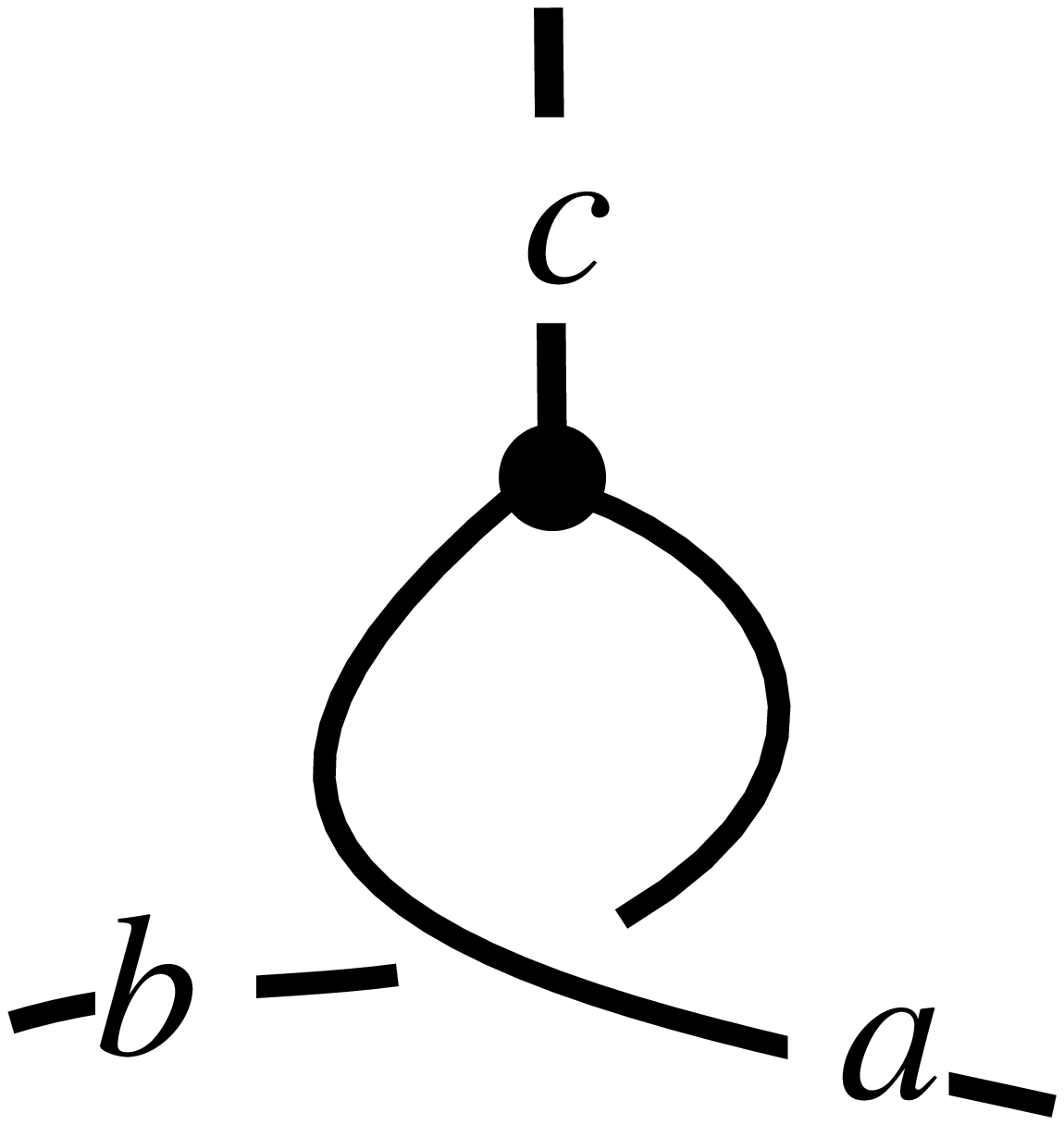} &= \lambda^{ab}_c \kd{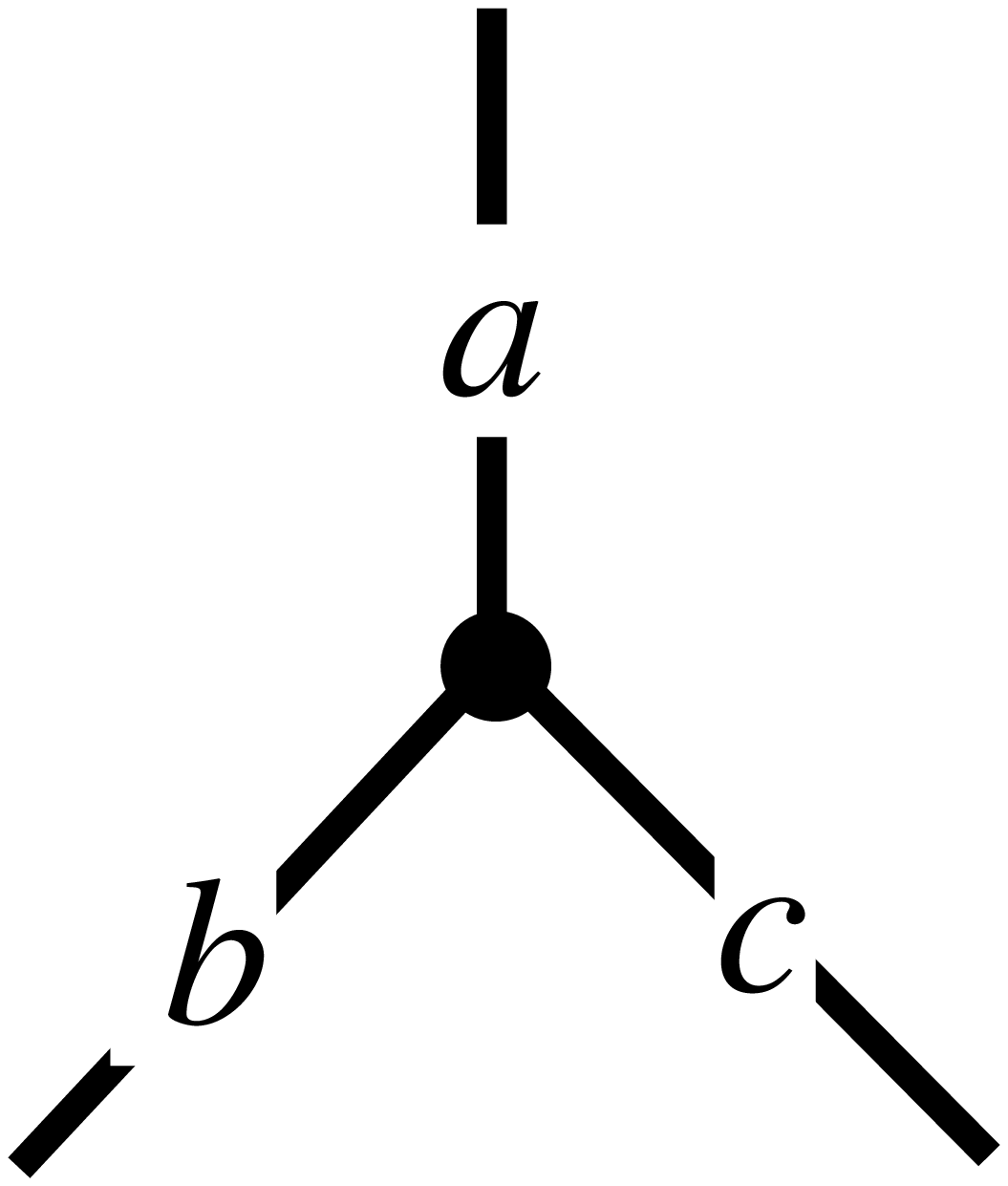}  \text{where $\lambda^{ab}_c$ 
is} \\
\lambda^{ab}_c &= (-1)^{[a^{2} + b^{2} - c^{2}]/2}.
\end{split}
\end{equation}



\begin{thebibliography}{999999}
	
\bibitem{penrose} Roger Penrose,
	``Angular momentum: An approach
	to combinatorial spacetime'' in {\em Quantum Theory and Beyond}
	T. Bastin, ed. (Cambridge University Press, Cambridge, 1971);
	``Combinatorial Quantum Theory and Quantized Directions'' in  
	{\it Advances in Twistor Theory}, Research Notes in Mathematics 37, 
	L. P. Hughston and R. S. Ward, eds. (Pitman, San Francisco, 1979) 
	pp. 301-307;
	``Theory of Quantized Directions,'' unpublished notes.

\bibitem{RSSpin} Carlo Rovelli and Lee Smolin,
"Spin Networks and Quantum Gravity,"
{\em Phys. Rev.} {\bf D 52} 5743-5759 (1995).

\bibitem{BaezSpin} John C. Baez, ``Spin networks in gauge theory,'' 
{\em Advances in Mathematics} {\bf 117} 253 - 272 (1996), Online Preprint 
Archive: {\em http://xxx.lanl.gov/abs/gr-qc/9411007}; ``Spin Networks in 
Nonperturbative Quantum Gravity,'' in {\em The Interface
of Knots and Physics},  Louis Kauffman, ed. 
(American Mathematical Society, Providence, Rhode 
Island, 1996), pp. 167 - 203, Online Preprint Archive:
{\em http://xxx.lanl.gov/abs/gr-qc/9504036}. 

\bibitem{mach} Ernst Mach, {\em Fichtes Zeitschrift f\"ur Philosophie}
{\bf 49} 227 (1866).  Cited in Lee Smolin in {\em Conceptual Problems
of Quantum Gravity}, A. Ashtekar and J. Stachel, eds. (Birkh\"auser,
Boston, 1991).

\bibitem{JM} John P. Moussouris, ``Quantum models as spacetime based 
	on recoupling theory,'' Oxford Ph.D. dissertation, unpublished (1983).
	
\bibitem{rsarea} 
Carlo Rovelli and Lee Smolin, ``Discreteness of area and volume in 
quantum gravity,'' {\em Nuc. Phys.} {\bf B 442}, 593-622 (1995).

\bibitem{RCgeometry} Roberto De Pietri and Carlo Rovelli, ``Geometry
eigenvalues and the scalar product from recoupling theory in loop
quantum gravity,'' {\em Phys. Rev. D} {\bf 54}(4), 2664-2690 (1996).

\bibitem{alarea} Abhay Ashtekar and Jerzy Lewandowski, ``Quantum 
Theory of Geometry I: Area operators,'' {\em Class.
Quant. Grav.} {\bf 14}, A55-A81 (1997).

\bibitem{flrarea} S. Fittelli, L. Lehner, C. Rovelli, ``The complete 
spectrum of the area from recoupling theory in loop quantum 
gravity,'' {\em Class. Quant. Grav}. {\bf 13}, 2921-2932 (1996).

\bibitem{R} K. Reidemeister, {\em Knotentheorie} (Chelsea Publishing Co.,
New York, 1948), original printing (Springer, Berlin, 1932). See also
Louis Kauffman, {\em Knots and Physics}, pp. 16.
	
\bibitem{knotsandphysics}
Louis H. Kauffman, {\em Knots and Physics}, Series on Knots and 
Everything - Vol. 1 (World Scientific, Singapore, 1991) pp. 125-130, 443-471.

\bibitem{KL}
Louis H. Kauffman and S\'ostenes L. Lins, {\em Temperley-Lieb 
Recoupling Theory and Invariants of 3-Manifolds}, Annals of 
Mathematics Studies N. 134, (Princeton University Press, Princeton, 
1994), pp. 1-100.

\bibitem{RS} Carlo Rovelli and Lee Smolin, ``Loop Representation of 
	quantum general relativity,'' {\em Nuc. Phys.} {\bf B 331}(1),
	80-152 (1990).
	
	
\bibitem{roberto}  R. DePietri, ``On the relation between the 
connection and the loop representation of quantum gravity,''
{\em Class. Quant. Grav.} {\bf 14} 53-70 (1990).

\bibitem{mess}
Albert Messiah,
{\em Quantum Mechanics},
Vol. 2
(John Wiley, New York, 1966).

\bibitem{jones}
H.F. Jones,
{\em Groups, Representations and Physics}
(Adam Hilger, Bristol, 1990).

\bibitem{AA} Abhay Ashtekar, ``New variables for classical and quantum
gravity,'' {\em Phys. Rev. Lett.} {\bf 57}(18), 2244-2247 (1986);
{\it New perspectives in canonical gravity} 
(Bibliopolis, Naples, 1988); { \it Lectures on non-perturbative canonical 
gravity}, Advanced Series in Astrophysics and Cosmology-Vol. 6
(World Scientific, Singapore, 1991).

\bibitem{AArev} Abhay Ashtekar, ``Quantum mechanics of Riemannian
	geometry,''
	{\em http://vishnu.nirvana.phys.psu.edu/riem$\_$qm/riem$\_$qm.html}.
	
\bibitem{CRrev} Carlo Rovelli, ``Loop Quantum Gravity,''
	{\em Living Reviews in Relativity}
	http://www.livingreviews.org/Articles/Volume1/1998-1rovelli;
	``Strings, Loops, and Others: A critical survey of the present 
	approaches to quantum gravity,'' in {\em Gravitation and Relativity: 
	At the turn of the Millennium}, Proceedings of the GR-15 Conference, 
	Naresh Dadhich and Jayant Narlikar, ed. (Inter-University Center 
	for Astronomy and Astrophysics, Pune, India, 1998), pp. 281 - 331,
	Online Preprint Archive: {\em http://xxx.lanl.gov/abs/gr-qc/9803024}.


\end{thebibliography}
\end{document}